\documentclass[aps,prl,longbibliography,floatfix,reprint]{revtex4-1}
\usepackage{amsmath}
\usepackage{graphicx}
\usepackage{amsfonts}
\usepackage{color}
\usepackage{placeins}

\begin{document}

\title{Phonon Boltzmann equation non-local in space and time: the partial failure of the generalized Fourier law}

\author{ Philip B. Allen }
\email{philip.allen@stonybrook.edu}
\affiliation{ Department of Physics and Astronomy,
              Stony Brook University, 
              Stony Brook, New York 11794-3800, USA }

\date{\today}

\begin{abstract}

The purpose of this note is to clarify the
solution of the non-local Peierls Boltzmann equation found by Hua and Lindsay (Phys. Rev. B {\bf 102}, 104310 (2020)).
They used methods of Cepellotti and Marzari.
The response function ``thermal distributor'' is discussed.
The new, ``non-Fourier'' term $\vec{B}$ [$\vec{J}_{\rm th}=-\kappa \vec{\nabla}T +\vec{B}]$ that occurs in
non-local situations, gives rise also to a new term in the thermal distributor.  


\end{abstract}

\maketitle

\section{Fourier law and nonlocal correction}

Transport in bulk systems, with currents small enough to linearize, is given by simple linear response functions,
such as the Fourier law for thermal conductivity, $\vec{J}_{\rm th}=-\kappa\vec{\nabla}T$, where the temperature
is close to a background temperature $T_0$, deviating by a small correction $\vec{r}\cdot\vec{\nabla}T$, with
$\vec{\nabla}T$ uniform in space.  However, in small systems, the current and
temperature need to be studied on distance scales less than the mean free paths of heat carriers.  A sensible assumption
is that a non-local Fourier law, $\vec{J}_{\rm F}(\vec{r})=-\int d\vec{r}^{ \ \prime} \kappa(\vec{r},\vec{r}^{ \ \prime})
\vec{\nabla}T(\vec{r}^{ \ \prime})$
should describe the linear response.  However, for non-uniform heat driving, a correction has been noticed recently.
An additional, non-Fourier term $\vec{J}_{\rm non-F}(\vec{r})$ appears.
This was found recently by Hua and Lindsay \cite{Hua2019,Hua2020}.  Hints of this can be found in earlier literature.
Boltzmann theory shows how this happens.  The derivation given below assumes heat carried by phonons in insulators.
Similar results are expected for heat carried by electrons in metals, and also for electrical conduction.
 The derivation includes time-dependent driving of phonon heat. 

\section{Non-local Peierls Boltzmann equation}

The phonon, or Peierls, Boltzmann equation \cite{Peierls1929,Ziman1960} (PBE), linearized for small deviations from equilibrium,
says that the phonon distribution $N_Q(\vec{r},t)$ evolves according to
\begin{eqnarray}
\frac{dN_Q}{dt}&=&\left(\frac{dN_Q}{dt}\right)_{\rm drift}
+\left(\frac{dN_Q}{dt}\right)_{\rm coll} +\left(\frac{dN_Q}{dt}\right)_{\rm ext}, \nonumber \\
&&  {\rm where} \ \left(\frac{dN_Q}{dt}\right)_{\rm drift}=-\vec{v}_Q \cdot\vec{\nabla}_{\vec{r}}N_Q,
\nonumber \\
&& {\rm and} \ \left(\frac{dN_Q}{dt}\right)_{\rm coll}=-\sum_{Q^\prime}R_{QQ^\prime}
\Delta N_{Q^\prime}(\vec{r},t), \nonumber \\
&& {\rm and} \ \left(\frac{dN_Q}{dt}\right)_{\rm ext}=\frac{1}{C(T_0)} \left(\frac{dn_Q}{dT}\right)_{T_0}P_Q(\vec{r},t). \nonumber \\
\label{eq:BE1}
\end{eqnarray}
The symbol $Q=(\vec{q},j)$ enumerates the $N=3nN_{\rm cell}$ eigenstates of the harmonic vibrational sustem.
The scattering operator $R_{QQ^\prime}$ is linearized for small deviations $\Delta N_Q$ from the local equilibrium
distribution, a Bose-Einstein distribution $n_Q$ at the local temperature $T(\vec{r},t)$;
$P_Q(\vec{r},t)$ is the rate per unit volume
of energy input into mode $Q$, and $T_0$ is the background temperature.
The specific heat $C(T_0)$ has units energy divided by temperature$\times$volume, and $P_Q$ has units power per
volume.  The external power input $P_Q$ causes mode $Q$ to increase in energy per unit time by an
amount which would increase its effective temperature by $d\Delta T_Q(\vec{r},t)/dt=P_Q(\vec{r},t)/C$.  
For simple driving, with $P_Q$ independent of $Q$, the modal temperature increase $\Delta T_Q$ is the
same for all modes $Q$.

Non-local Boltzmann theories require a definition of local temperature; the one that works in Boltzmann theory is 
that the local energy density $u(\vec{r},t)$ is $(1/V)\sum_Q \hbar\omega_Q n_Q(T(\vec{r},t))$. 
This means that there is no
energy in the deviation $\Delta N_Q\equiv N_Q(\vec{r},t)-n_Q(T(\vec{r},t))$.  The sum $(1/V)\sum_Q \hbar\omega_Q 
\Delta N_Q =0$ defines $T(\vec{r},t)$.

\section{Scattering and Temperature}

Each scattering event conserves phonon energy: $\sum_Q\hbar\omega_Q (d\Delta N_Q/dt)_{\rm scatt}=0$,
whether linearized in $\Delta N_Q$ or not.  In linear approximation, this is equivalent to the sum rule
$\sum_Q \hbar\omega_Q R_{QQ^\prime}=0$.  The operator $\hat{R}$ (where $R_{QQ^\prime}=
\langle Q|\hat{R}|Q^\prime\rangle$) has a left null eigenvector $\langle \hbar\omega|\hat{R}=0$, where
$\hbar\omega_Q = \langle \hbar\omega|Q\rangle$.  The vector $|\cdot\rangle$ lies in the
space of harmonic eigenstates.  The states $|Q\rangle$ are a complete orthonormal basis.  In the
coordinate space enumerated by unit cell $\ell$ at $\vec{R}_\ell$, and atom number $n$, and Cartesian coordinate $i$,
\begin{equation}
\langle \ell n i |Q\rangle = \frac{1}{\sqrt{N_{\rm cells}}} e^{i\vec{q}\cdot\vec{R}_\ell} \epsilon_{Q}(n,i),
\label{eq:}
\end{equation}
where the polarization vector $\hat{\epsilon}_Q$ is normalized, $\sum_{n,i} |\epsilon_Q(n,i)|^2=1$.
But $\hat{R}$ is not symmetric.  The corresponding
null right eigenvector is $|n(n+1) \hbar\omega\rangle$.   This is equivalent to the statement that if the
distribution $N_Q=n_Q(T(\vec{r},t)) +\Delta N_Q(\vec{r},t)$ consists of shifting the occupancy of every mode $Q$
away from local equilbrium $T(\vec{r},t)$ by
$\Delta N_Q=(dn_Q/dT)\Delta T(\vec{r},t)$ with the same $\Delta T(\vec{r},t)$ for each mode, 
the shifted distribution is already locally thermalized and doesn't experience net scattering.

A symmetrized scattering operator $\hat{\Omega}$ is \cite{Simons1960,Krumhansl1965}
\begin{eqnarray}
\hat{\Omega}&=&[\hat{n}(\hat{n}+1)]^{-1/2} \hat{R} [\hat{n}(\hat{n}+1)]^{1/2} \nonumber \\
\Omega_{QQ^\prime}&=&[n_Q(n_Q+1)]^{-1/2} R_{QQ^\prime} [n_{Q^\prime}(n_{Q^\prime}+1)]^{1/2}.
\label{eq:Omega}
\end{eqnarray}
The operator $\hat{n}$ is defined as $\langle Q|n|Q^\prime\rangle=n_Q \delta_{QQ^\prime}$.
The Bose factors $n_Q$ in Eq. \ref{eq:Omega} are taken at the background temperature $T_0$, not the 
local equilibrium temperature $T(\vec{r},t)$.  Similarly the operator $\hat{n}$ is defined at
$T=T_0$ unless explicitly written as $\hat{n}(\vec{r},t)$.
The Boltzmann Eq. \ref{eq:BE1} is now
\begin{eqnarray}
\frac{dN(\vec{r},t)\rangle}{dt} &=& -\hat{\vec{v}}\cdot\vec{\nabla}_{\vec{r}} |N(\vec{r},t)\rangle \nonumber \\
&-&\sqrt{\hat{n}(\hat{n}+1)}\hat{\Omega} \frac{1}{\sqrt{\hat{n}(\hat{n}+1)}} |\Delta N(\vec{r},t)\rangle \nonumber \\
&+&\hat{n}(\hat{n}+1)\frac{\hbar\hat{\omega}}{Ck_B T^2}|P(\vec{r},t)\rangle,
\label{eq:BE2}
\end{eqnarray}
where $\langle Q|P\rangle=P_Q$  The operators $\hat{\vec{v}}$ and $\hat{\omega}$ are diagonal in $Q$-representation, $\langle Q|\hat{\vec{v}}|Q^\prime\rangle
=\vec{v}_Q \delta_{QQ^\prime}$ and $\langle Q|\hat{\omega}|Q^\prime\rangle =\omega_Q \delta_{QQ^\prime}$.

Now define from $\Delta N_Q$ a deviation function $\Phi_Q$,
\begin{eqnarray}
\Delta N_Q &=& N_Q - n_Q(\vec{r},t) \equiv  \sqrt{n_Q(n_Q+1)}\Phi_Q(\vec{r},t) \nonumber \\
|\Delta N\rangle&=& |N\rangle-|n(\vec{r},t)\rangle \equiv \sqrt{\hat{n}(\hat{n}+1)}|\Phi(\vec{r},t)\rangle
\label{eq:}
\end{eqnarray}
Then Eq. \ref{eq:BE2} becomes
\begin{equation}
\frac{\partial |\Phi(\vec{r},t)\rangle}{\partial t}+\hat{\vec{v}}\cdot\vec{\nabla}_{\vec{r}}|\Phi(\vec{r},t)\rangle 
+ \hat{\Omega}|\Phi(\vec{r},t)\rangle =|A(\vec{r},t)\rangle,
\label{eq:BE3}
\end{equation}
\begin{eqnarray}
|A(\vec{r},t)\rangle&=&-[\hat{n}(\hat{n}+1)]^{-1/2} \left[\left(\frac{\partial}{\partial t}+\hat{\vec{v}}
\cdot\vec{\nabla}_{\vec{r}}\right) \Delta T(\vec{r},t) \right] \left| \frac{dn}{dT}\right\rangle \nonumber \\
&+&[\hat{n}(\hat{n}+1)]^{1/2} \frac{\hbar\hat{\omega}}{Ck_B T^2} |P(\vec{r},t)\rangle,
\label{eq:A1}
\end{eqnarray}
The inhomogeneous part $|A\rangle$ has been separated.  It drives the deviation $|\Phi\rangle$
from local equilibrium, $|\Phi\rangle=0$.
Because of linearization, the equation is simplified by a Fourier transform,
\begin{eqnarray}
|X(\vec{k},\omega)\rangle&=&\frac{1}{V}\int^V d\vec{r} \int_{-\infty}^\infty dt |X(\vec{r},t)\rangle 
\exp[-i(\vec{k}\cdot\vec{r}-\omega t)] \nonumber \\
|X(\vec{r},t)\rangle&=&\sum_{\vec{k}} \int_{-\infty}^\infty d\omega |X(\vec{k},\omega)\rangle \exp[i(\vec{k}\cdot\vec{r}-\omega t)]
\label{eq:}
\end{eqnarray}
The different Fourier components are 
not coupled, and are treated one at a time, as if $|X(\vec{r},t)\rangle=|X(\vec{k},\omega)\rangle 
\exp[i(\vec{k}\cdot\vec{r}-\omega t)]$.
The symbol $\omega$ is the external frequency, not to be confused with $\hat{\omega}$ which is the operator version of
the phonon frequency $\omega_Q$.
Equations \ref{eq:BE3} and \ref{eq:A1} become
\begin{equation}
\left[ \hat{\Omega} +i(\vec{k}\cdot\hat{\vec{v}}-\omega\hat{1}) \right] |\Phi\rangle = |A(\vec{k},\omega)\rangle
\label{eq:BEkw}
\end{equation}
\begin{equation}
|A\rangle=\frac{\sqrt{\hat{n}(\hat{n}+1)}}{k_B T^2}
\left[- i(\vec{k}\cdot\hat{\vec{v}}-\omega\hat{1})\Delta T |\hbar\omega\rangle +\frac{\hbar\hat{\omega}}{C} |P\rangle \right]
 \label{eq:Avec}
\end{equation}

Mode space has so far been described in $Q$-representation by harmonic eigenstates $|Q\rangle$.
It is convenient to also use the ``relaxon"-representation  $|\alpha\rangle$ of eigenstates 
\cite{Guyer1966,Maris1969,Cepellotti2016} of $\hat{\Omega}$.
\begin{equation}
\hat{\Omega} |\alpha\rangle = \gamma_\alpha |\alpha\rangle
\label{eq:eigW}
\end{equation}
where
\begin{equation}
\langle\alpha|\beta\rangle=\delta_{\alpha\beta} \ \ {\rm and} \ \ \sum_\alpha |\alpha\rangle\langle\alpha|=\hat{1}.
\label{eq:orth}
\end{equation}
The eigenvalues $\gamma_\alpha$ are relaxation rates, $\gamma_\alpha=1/\tau_\alpha$.
In this basis, Eq. \ref{eq:BEkw} has the form
\begin{equation}
\sum_\beta\left[ \gamma_\alpha \delta_{\alpha\beta} +i(\vec{k}\cdot\vec{v}_{\alpha\beta} -\omega\delta_{\alpha\beta}) \right]
\Phi_\beta = A_\alpha,
\label{eq:BErr}
\end{equation}
where $\Phi_\beta = \langle \beta |\Phi\rangle$ and $A_\alpha=\langle\alpha |A\rangle$.
There are $N$ modes $Q$ ($N=3 N_{\rm cells} n_{\rm at}$ where $n_{\rm at}$ is the number of atoms in the unit cell),
and Eq. \ref{eq:BErr} gives $N$ equations for the $N$ unknown components
$\Phi_\alpha$ of the deviation function.  But there are two fields ($|\Delta T\rangle$ and $|P\rangle$)
driving the distribution out of equilibrium, of which one (typically $|\Delta T\rangle$) is unknown.  
An extra equation is needed.  That equation is the definition of local temperature.

There is one null eigenvector, $|\alpha=0\rangle \equiv |0\rangle$ with eigenvalue $\gamma_0=0$.  
The vectors $\langle 0|$ and $|0\rangle$ deviate only by 
factors $[\hat{n}(\hat{n}+1)]^{\pm 1/2}$ from the null left $(\langle\hbar\omega|)$ and right 
$|dn/dT\rangle$ eigenvectors of $\hat{R}$.  In $Q$-representation, the null eigenvector is
\begin{equation}
\langle Q|0\rangle = \sqrt{n_Q(n_Q + 1)} \frac{\hbar\omega_Q}{\sqrt{CVk_B T^2}}
\label{eq:null}
\end{equation}
 The factor $ 1/ \sqrt{VCk_B T^2}$ normalizes the state, $\langle 0|0 \rangle=1$. 
The definition of temperature takes the form
\begin{eqnarray}
\frac{1}{V}\sum_Q && \hbar\omega_Q \Delta N_Q =0=\frac{1}{V}\sum_Q \hbar\omega_Q\sqrt{n_Q(n_Q+1)}\Phi_Q \nonumber \\
&&=\sqrt{\frac{Ck_B T^2}{V}} \langle 0|\Phi\rangle = \sqrt{\frac{Ck_B T^2}{V}} \Phi_0 =  0.
\label{eq:}
\end{eqnarray}
This shows that there are actually only $N-1$ unknown parts of $|\Phi\rangle$, because the $\alpha=0$ 
component, $\langle 0|\Phi\rangle=\Phi_0$ must be zero by the definition of local temperature.

Now we can rewrite the formula for $|A\rangle$ using Eqs. \ref{eq:Avec} and \ref{eq:null},
\begin{equation}
|A\rangle=\sqrt{\frac{V}{Ck_B T^2}} \left[- i(\vec{k}\cdot\hat{\vec{v}}-\omega\hat{1})
 C\Delta T(\vec{k},\omega) + \hat{P}(\vec{k},\omega) \right] |0 \rangle
\label{eq:}
\end{equation}
where $\langle Q|\hat{P}|Q^\prime\rangle =P_Q \delta_{QQ^\prime}$.
Now look at the $\alpha=0$ component of Eq. \ref{eq:BErr},
\begin{equation}
i\vec{k}\cdot\sum_\beta^{\ne 0} \vec{v}_{0\beta}\Phi_\beta = A_0
\label{eq:0comp}
\end{equation}
The left hand side of Eq. \ref{eq:0comp} is
\begin{equation}
\langle 0   | i\vec{k}\cdot\vec{v}|\Phi\rangle = \sqrt{\frac{V}{Ck_B T^2}} i\vec{k}\cdot\vec{J}
\label{eq:}
\end{equation}
where $\vec{J}(\vec{k},\omega)$ is the energy (or heat) current density, $(1/V)\sum_Q \hbar\omega_Q \vec{v}_Q 
[N_Q(\vec{k},\omega)-n_Q(\vec{k},\omega)]$.  The right hand side of Eq. \ref{eq:0comp} is
\begin{equation}
\langle 0 |A\rangle=\sqrt{\frac{V}{Ck_B T^2}}\left( i\omega C\Delta T + \bar{P} \right),
\label{eq:}
\end{equation}
where $\bar{P}=\langle 0|\hat{P}|0 \rangle$, or
\begin{equation}
\bar{P}\equiv \sum_Q \frac{C_Q}{C} P_Q \ \ {\rm and} \ \ C_Q=\frac{1}{V} \hbar\omega_Q \frac{dn_Q}{dT},
\label{eq:}
\end{equation}
and $C=\sum_Q C_Q$ is the total specific heat.  If $P_Q$ is independent of $Q$, then $P_Q=\bar{P}$ for all modes $Q$.
Thus the $\alpha=0$ part of the Boltzmann equation expresses energy conservation.  In $(\vec{r},t)$ representation,
\begin{equation}
\vec{\nabla}_{\vec{r}} \cdot \vec{J} = -\frac{\partial u}{\partial t} +\bar{P}.
\label{eq:Econs}
\end{equation}
where $\partial u/\partial t=C \partial \Delta T/\partial t \rightarrow -i\omega C \Delta T$.

We can now ``reinterpret'' Eq. \ref{eq:BErr} as a system of $N-1$ linear equations (the sum over $\beta$ does
not include $\beta=0$ because $\Phi_0 = 0$).  The $(N-1)\times(N-1)$ matrix $\Omega_{\alpha\beta}=\gamma_\alpha
\delta_{\alpha\beta}$ is now positive-definite.  We still need to invert the non-Hermitean matrix on the left side
of Eq. \ref{eq:BErr} in order to solve for $|\Phi\rangle$.



\section{The Solution}
\label{sec:FT}

The method of solution is given by Hua and Lindsay \cite{Hua2020}.
A earlier version is in a paper by Cepellotti and Marzari \cite{Cepellotti2017b}.
Rewrite Eq. \ref{eq:BErr} by rescaling the distribution function $\Phi_\alpha$ and driving term $A_\alpha$:
\begin{equation}
\Psi_\beta\equiv\gamma_\beta^{1/2} \Phi_\beta \  {\rm and} \  B_\alpha\equiv\gamma_\alpha^{-1/2}A_\alpha.
\label{eq:Psidef}
\end{equation}
Equation \ref{eq:BErr} then becomes $[\hat{1} + i\hat{\Gamma}]|\Psi\rangle=|B\rangle$, or
\begin{eqnarray}
&&\sum_\beta \left[\delta_{\alpha\beta}+ i\Gamma_{\alpha\beta} \right] \Psi_\beta = B_\alpha  \nonumber \\
\Gamma_{\alpha\beta}&=&\vec{k}\cdot \left( \gamma_\alpha^{-1/2} \vec{v}_{\alpha\beta} \gamma_\beta^{-1/2} \right)
-\gamma_\alpha^{-1} \omega \delta_{\alpha\beta}
\label{eq:BEg}
\end{eqnarray}
The matrix $\hat{\Gamma}$ is real-symmetric, so it has real eigenvalues, $\lambda_m$:
\begin{equation}
\hat{\Gamma}(\vec{k},\omega)|m\rangle = \lambda_m(\vec{k},\omega) |m\rangle.
\label{eq:}
\end{equation}
In this basis, Eq. \ref{eq:BEg} is $(1+i\lambda_m) \Psi_m = B_m$, where $\Psi_m=\langle m|\Psi\rangle$, {\it etc.}
Then the distribution function in the relaxon basis (the eigenbasis of $\hat{\Omega}$) is
\begin{eqnarray}
&&\Psi_\alpha=\sum_\beta S_{\alpha\beta} B_\beta, \ {\rm or} \ \ \Phi_\alpha=\sum_\beta
\left(\gamma_\alpha^{-1/2} S_{\alpha\beta} \gamma_\beta^{-1/2} \right) A_\beta,
\nonumber \\ 
&&{\rm where} \ \ \langle \alpha |\hat{S}| \beta\rangle = S_{\alpha\beta}=\sum_m \langle \alpha |m \rangle
\frac{1}{1+i\lambda_m(\vec{k},\omega)} \langle m |\beta \rangle. \nonumber \\
\label{eq:PsiS}
\end{eqnarray}
This is the desired solution.  In the spatially homogeneous ($\vec{k}=0$) and static ($\omega=0$) case,
$\hat{\Gamma}=0$ and $\lambda_m=0$, so $S_{\alpha\beta}=\delta_{\alpha\beta}$ and 
the bulk solution $\Phi_\alpha=\gamma_\alpha^{-1}A_\alpha$ is recovered.

\section{Heat current}

The operator $\hat{\Omega}$ is positive if we exclude the null
space.  The operator $\hat{S}=(\hat{1}+i\hat{\Gamma})^{-1}$ is defined in this ``positive'' or $p$-space.
It is convenient to define another operator in the same $p$-space,
\begin{eqnarray}
\hat{W}&\equiv& \hat{\Omega}^{-1/2} \hat{S} \hat{\Omega}^{-1/2} \nonumber \\
W_{\alpha\beta}&=&\gamma_\alpha^{-1/2} S_{\alpha\beta} \gamma_\beta^{-1/2} 
\label{eq:}
\end{eqnarray}
The solution of the Boltzmann equation is then
\begin{equation}
|\Phi\rangle = \hat{W} |A\rangle_p
\label{eq:}
\end{equation}
where $|A\rangle_p$ is the part of the driving term that is orthogonal to $|0\rangle$ (and thus lies in $p$-space),
\begin{equation}
|A\rangle_p = \sqrt{\frac{V}{Ck_B T^2}} \left[ -i\vec{k}\cdot\hat{\vec{v}} C\Delta T +(\hat{P}-\bar{P}\hat{1}) \right] |0\rangle.
\label{eq:}
\end{equation}
As required, the inner product
$\langle 0|A\rangle_p$ vanishes, because it has a term proportional to  $\langle 0|\hat{\vec{v}}|0\rangle$ 
which is zero because $\hat{\vec{v}}$ is an odd operator, and a term proportional to $\langle 0|(\hat{P}-\bar{P}\hat{1})|0\rangle$
that is zero because $\langle 0|\hat{P}|0 \rangle \equiv \bar{P}$.
From $|\Phi\rangle$ we get the heat current density $\vec{J}$,
\begin{equation}
\vec{J}=\sqrt{\frac{Ck_B T^2}{V}} \langle 0|\hat{\vec{v}} |\Phi\rangle = \vec{J}_{\rm F} + \vec{J}_{\rm non-F}.
\label{eq:}
\end{equation}
The ``generalized Fourier'' component $\vec{J}_{\rm F}$ is
\begin{equation}
\vec{J}_{\rm F}(\vec{k},\omega)= -\kappa(\vec{k},\omega)\cdot(i\vec{k}) \Delta T(\vec{k},\omega),
\label{eq:}
\end{equation}
\begin{equation}
\kappa(\vec{k},\omega)=C \  \langle 0 |\hat{\vec{v}} \ \hat{W}(\vec{k},\omega) \ \hat{\vec{v}} |0\rangle.
\label{eq:}
\end{equation}
The ``non-Fourier'' term is
\begin{equation}
\vec{J}_{\rm non-F}(\vec{k},\omega)= 
\langle 0| \vec{\hat{v}} \hat{W} [\hat{P}(\vec{k},\omega)-\bar{P}(\vec{k},\omega)\hat{1}] |0\rangle.
\label{eq:}
\end{equation}
This extra, non-Fourier, component of the current was found by Hua {\it et al.} \cite{Hua2019}
in an RTA treatment.  A version is in the paper by Mahan and Claro \cite{Mahan1988}.
In their version, there was no external insertion $P$ except {\it via} boundary conditions.
 It was rederived by Hua and Lindsay \cite{Hua2020} in a more complete treatment (not using RTA).
 Reference \onlinecite{Hua2019} gives additional parts of the non-Fourier
 term that arise from boundary conditions, but solved only in RTA.
Boundary terms do not appear here; the present derivation, like ref. \onlinecite{Hua2020},
 assumes an infinite homogeneous sample.
 
In a d.c. situation ($\omega=0$), the insertion term $\hat{P}(\vec{r})$ should have a
zero spatial average: $P(\vec{k}=0,\omega=0)=0$.  Otherwise, the sample experiences net heating or cooling.
The spatially homogeneous ($\vec{k}=0$) part of the ``non-Fourier'' current is zero.  Also,
when $\vec{k}=0$ and $\omega=0$, $\hat{W}(0,0)=\hat{\Omega}^{-1}$ and the bulk
static thermal conductivity is recovered,
\begin{equation}
\kappa(\vec{k}=0,\omega=0)=C \  \langle 0 |\hat{\vec{v}} \ \hat{\Omega}^{-1} \ \hat{\vec{v}} |0\rangle.
\label{eq:}
\end{equation}

\section{Thermal distributor}

The local temperature distribution $T(\vec{r},t)$ can be found from energy conservation (Eq. \ref{eq:Econs}),
\begin{equation}
\Delta T(\vec{k},\omega) =\frac{1}{i\omega C} \left[
i\vec{k}\cdot \left(\vec{J}_{\rm F}(\vec{k},\omega)+\vec{J}_{\rm non-F}(\vec{k},\omega)\right)- \bar{P}(\vec{k},\omega)\right].
\label{eq:}
\end{equation}
If $P_Q(\vec{k},\omega)$ is independent of $Q$ ({\it i.e.} $\hat{P}=\bar{P}$), the non-Fourier current is zero,
and the local temperature variation is given by a simple nonlocal response function, the ``thermal distributor'' $\Theta(\vec{k},\omega)$,
\begin{equation}
\Delta T(\vec{k},\omega)=\Theta(\vec{k},\omega)\bar{P}(\vec{k},\omega)
\label{eq:thd}
\end{equation}
where
\begin{equation}
\Theta(\vec{k},\omega)=\frac{1}{\vec{k}\cdot \kappa(\vec{k},\omega)\cdot\vec{k} - i\omega C}
\label{eq:thdd}
\end{equation}
This is Eq. 38 of ref. \onlinecite{Hua2020}.
The function $\Theta$ was defined in ref. \onlinecite{Allen2018}.
The name ``thermal distributor,''  has been changed from the previous name ``thermal susceptibility,'' and a factor of C
has been removed from the definition.
When $\hat{P} \ne \bar{P}$, there is another term,
\begin{equation}
\Delta T(\vec{k},\omega) = \Theta(\vec{k},\omega) \bar{P}(\vec{k},\omega) - \frac{i\vec{k}\cdot \vec{J}_{\rm non-F}(\vec{k},\omega)}
{\vec{k}\cdot \kappa(\vec{k},\omega)\cdot\vec{k} - i\omega C}
\label{eq:}
\end{equation}

Griffin \cite{Griffin1965} makes an interesting argument that may be related to thermal distributors. 
Guyer and Krumhansl \cite{Guyer1966} noticed that Griffin's argument is related to the mean-field method relating
non-interacting to interacting
electrical susceptibility. 


%
\section{appendix: terminology}

Various names are given to the operator $-\vec{v}_Q\cdot\vec{\nabla}_{\vec{r}}$:
(a) ``drift operator'', (b) ``advection operator", (c) ``convection operator'', or (d) ``diffusion operator".
Sometimes the combination $\partial /\partial t +\vec{v}_Q\cdot\vec{\nabla}_{\vec{r}}$  is called
the ``drifting operator'' . 
The process these words are describing is that the occupancy $N_Q(\vec{r},t+\Delta t)$, in a 
completely ballistic system, is the same as $N_Q(\vec{r}-\vec{v}_Q \Delta t,t)$, so that 
\begin{eqnarray}
 \left(\frac{dN_Q}{dt}\right)_{\rm drift}&=&
\lim_{\Delta t \rightarrow 0} \left[ \frac{N_Q(\vec{r},t+\Delta t)-N_Q(\vec{r}, t)}{\Delta t} \right]_{\rm ballistic} \nonumber \\
&=& -\vec{v}_Q \cdot \vec{\nabla}_{\vec{r}}N_Q \nonumber.
\label{eq:}
\end{eqnarray}
The term I use, ``drift operator'', makes sense; ``diffusion operator" does not.


\bibliography{BibJuly}

\end{document}